\documentclass[11pt, aps, prd, preprint,amsmath,amssymb]{revtex4-1}
\usepackage{graphicx}

\begin{document}
\title{Acceleration without Horizons}
\author{Alaric Doria and Gerardo Mu{\~n}oz}
\email{alaricdoria@mail.fresnostate.edu, gerardom@csufresno.edu}
\affiliation{Department of Physics, California State University, Fresno, Fresno, CA 93740-8031}

\date{\today}

\begin{abstract}
We derive the metric of an accelerating observer moving with non-constant proper acceleration in flat spacetime. With the exception of a limiting case representing a Rindler observer, there are no horizons. In our solution, observers can accelerate to any desired terminal speed $v_{\infty} < c$. The motion of the accelerating observer is completely determined by the distance of closest approach and terminal velocity or, equivalently, by an acceleration parameter and terminal velocity.
\end{abstract}
\maketitle
\newcommand{\half}{\tfrac{1}{2}}
\newcommand{\quarter}{\tfrac{1}{4}}
\newcommand{\third}{\tfrac{1}{3}}
\newcommand{\twothirds}{\tfrac{2}{3}}

\section{Introduction}

A complete description of accelerating observers in flat spacetime is lacking in the literature. Observers moving with constant proper acceleration (Rindler observers \cite{Rindler}) are the most widely used description of such behavior so far, especially in discussions of the thermal properties of vacuum states in quantum field theories \cite{Crispino}. In this paper we derive the trajectory and the metric of an observer moving with non-constant proper acceleration in two-dimensional flat spacetime.

It is clear that Rindler observers are unphysical. They accelerate with constant proper acceleration for all time, reaching speeds arbitrarily close to the speed of light. For this reason, Rindler observers see the light cone as an event horizon.

These horizons partition the spacetime into four regions \cite{Rindler}: the left and right Rindler wedges, which we characterize by a metric $ds^2=X^2 dT^2 - dX^2$, and the Kasner-Milne \cite{Kasner, Milne, KM} universe, which we characterize by a metric $ds^2 = dT^2 - T^2dX^2$. These regions are bounded by the lines in flat spacetime at 45 degrees to the $x$- and $t$-axes (in units of $c=1$). Observers in the Rindler wedges are disconnected from some of the other wedges and thus do not have complete access to the full spacetime. For instance, an observer in the right Rindler wedge can receive signals from the past of the Kasner-Milne universe (lower wedge) and send signals to the future Kasner-Milne  universe (upper wedge), but is completely disconnected from the left Rindler wedge. This is because observers in the Rindler wedges see event horizons corresponding to the light cone portions in their quadrants of the spacetime. Minkowski (inertial) observers never experience this; an event occuring in flat spacetime is always accessible to the Minkowski observer provided the observer waits sufficiently long times. 

\section{Derivation of the Metric}

We begin with the flat spacetime metric
\begin{equation}
ds^2 = dt^2 - dx^2
\end{equation}
and change coordinates to the accelerating frame. The most general coordinate transformation is $t = h(\xi,\zeta)$ and $x = f(\xi,\zeta)$. These coordinate transformations lead to the following metric:
\begin{equation}
ds^2 = (\dot{h}^2 - \dot{f}^2)d\xi^2 + 2(\dot{h}h^\prime - \dot{f}f^\prime)d\xi d\zeta + (h^{\prime 2} - f^{\prime 2})d\zeta^2
\end{equation}

Here primes indicate derivatives with respect to $\zeta$ and dots indicate derivatives with respect to $\xi$. We attempt to eliminate the cross term $d\xi d\zeta$. We choose $\dot{h}h^\prime =\dot{f}f^\prime$ and solve the partial differential equation. We assume a solution that is separable, $t = a(\xi)b(\zeta)$ and $x = c(\xi)d(\zeta)$. It follows that
\begin{equation}
\frac{\dot{a}a}{\dot{c}c} = \frac{dd^\prime}{bb^\prime}
\end{equation}
Each side of the equation depends only on $\xi$ or $\zeta$ and therefore must be equal to a constant. Integrating both equations, we obtain relationships between the functions $a$ and $c$ as well as $d$ and $b$. 
\begin{equation}
\label{AB}
\begin{cases}
a^2 - k^2 c^2 = A^2 \\
d^2 - k^2 b^2 = B^2
\end{cases}
\end{equation}

The general solutions to the equations in (\ref{AB}) come in four cases. We consider the trivial case first, $A = B = 0$. This leads to functions $a=\pm kc$ and $d=\pm kb$ and forces $ds^2 = 0$. 

We take cases two and three to be $A^2=0$ and $B^2\not = 0$, and vice versa. First we take $A^2 = 0$; this reduces the equations in (\ref{AB}) to $a =\pm k c$ and $(d/B)^2 - (kb/B)^2 = 1$. The general solution to this equation is $d = B \cosh\phi(\zeta)$ and $b = Bk^{-1} \sinh\phi(\zeta)$. Thus, our coordinate transformations look like $t = B c(\xi) \sinh\phi(\zeta)$ and $x = B c(\xi) \cosh\phi(\zeta)$, and we see that case two corresponds to the left and right Rindler wedges. With coordinate choices $X = Bc(\xi)$ and $T=\phi(\zeta)$ the metric becomes
\begin{equation}
\label{R}
ds^2 = X^2 dT^2 - dX^2
\end{equation}

For case three, the condition $A^2\not=0$ and $B^2=0$ implies $t=Ab(\zeta)  \cosh\theta(\xi)$ and $x=A b(\zeta) \sinh\theta(\xi)$, which describe the Kasner-Milne universe. The coordinate choices $X=\theta(\xi)$ and $T = Ab(\zeta)$ now yield the metric
\begin{equation}
\label{K}
ds^2 = dT^2 - T^2 dX^2
\end{equation}
Observers in the Kasner-Milne wedges move with constant velocity; the distance between two observers decreases in the lower wedge and increases in the upper wedge, leading some to refer to the upper and lower wedges as the expanding and contracting universe, respectively \cite{Crispino}.

The fourth -- and most interesting -- case is $A^2\not=0$ and $B^2\not=0$. Then, the solutions to equations (\ref{AB}) with the condition that $A$ and $B$ are both non-zero become $a = A \cosh\phi(\xi)$, $c = A k^{-1} \sinh\phi(\xi)$, and also $d = B \cosh\theta(\zeta)$, $b = B k^{-1} \sinh\theta(\zeta)$. With the appropriate choice of $X$ and $T$ it follows that 
\begin{equation}
\label{traject}
t = \alpha \cosh\left( \frac{X}{\alpha} \right)\sinh\left( \frac{T}{\alpha} \right) \quad , \quad x = \alpha \sinh\left( \frac{X}{\alpha} \right)\cosh\left( \frac{T}{\alpha} \right)
\end{equation}
where $\alpha = AB k^{-1}$. The metric of the accelerating observer is therefore
\begin{equation}
\label{metric}
ds^2 = \cosh\left(\frac{X+T}{\alpha}\right)\cosh\left(\frac{X-T}{\alpha}\right)(dT^2 - dX^2)
\end{equation}

In Figure 1 we show a spacetime diagram with lines of constant $T$ and constant $X$. It is clear from this figure that the accelerating observer sees no event horizons. The trajectories are hyperbolic but they are not bounded by the light cones.

\section{Kinematics}

We select a fixed position $X_0$ in the instantaneous rest frame of the accelerating observer and calculate the velocity.
\begin{equation}
\frac{dx}{dt} = \tanh\left(\frac{X_0}{\alpha}\right)\tanh\left(\frac{T}{\alpha}\right) = v_\infty \tanh\left( \frac{T}{\alpha} \right)
\end{equation}

The speed of the accelerating observer ranges from 0 to 1, the speed of light, and we may adjust this speed by varying either the position of the accelerating observer in his coordinate system or the parameter $\alpha$. We define the terminal velocity $v_\infty \equiv \tanh(X_0/\alpha)$ for convenience and physical insight, and denote by $\gamma_\infty$ the relativistic gamma factor associated with this terminal velocity, so that $\gamma_\infty = \cosh(X_0/\alpha)$. We then calculate the acceleration $a = d^2 x / dt^2$. 
\begin{equation}
a = \frac{v_\infty}{\alpha} \frac{1}{\cosh\left( \frac{X_0}{\alpha} \right)\cosh^3\left( \frac{T}{\alpha} \right)} = \frac{v_\infty}{\alpha\gamma_\infty} \frac{1}{\cosh^3\left( \frac{T}{\alpha} \right)}
\end{equation}
The transformation law to obtain the proper acceleration is $a_{pr} = \gamma^3 a$.
\begin{equation}
a_{pr} = \frac{v_\infty}{\alpha\gamma_\infty}\left[1+\gamma_\infty^{-2}\sinh^2\left( \frac{T}{\alpha} \right)\right]^{-3/2}
\end{equation}

For completeness, we note from (\ref{metric}) or from the equivalent form
\begin{equation}
\label{metric2}
ds^2 = \left[ \cosh^2\left(\frac{T}{\alpha}\right) + \sinh^2\left(\frac{X}{\alpha}\right) \right] (dT^2 - dX^2)
\end{equation}
that the relationship between the time $T$ and the proper time $\tau$ of the observer at $X_0$ is
\begin{equation}
dT = \left[\gamma_\infty^{2} + \sinh^2\left( \frac{T}{\alpha} \right)\right]^{-1/2} d \tau
\end{equation}
Unlike in the Rindler case, $dT$ and $d \tau$ are related here by a time-dependent factor.

We have yet to identify the meaning of the parameter $\alpha$. To determine what $\alpha$ is physically, we need to analyze the trajectory $x(T)$ which we obtain from the original coordinate transformations (\ref{traject}). 
\begin{equation}
x(T) = \alpha \sinh\left( \frac{X_0}{\alpha} \right) \left[ 1 + \sinh^2\left( \frac{T}{\alpha} \right) \right]^{1/2}
\end{equation}
We also see from (\ref{traject}) that $t = 0 \Rightarrow T = 0$, and this is the exact condition we need to determine $\alpha$.
\begin{equation}
x_0 = \alpha \sinh\left( \frac{X_0}{\alpha} \right) = \alpha\gamma_\infty v_\infty
\end{equation}
Therefore, $\alpha=x_0 / v_\infty\gamma_\infty$ is determined by the distance of closest approach $x_0$ and the initial velocity $v_\infty$. Alternatively, we could also fix $\alpha$ by specifying the acceleration at $T=t=0$ which leads to $\alpha = v_\infty / \gamma_\infty a_{0}$, where $a_0 = a(0) = a_{pr}(0)$. In contrast to Rindler, where observers require infinite accelerations \cite{Rindler}, $X_0 = 0$ corresponds to stationary observers.

\section{Minkowski Coordinates}

We now return to our original kinematic equations and replace $\alpha$ with physical parameters and restore Minkowski time. This allows us to make comparisons with what we know about the equations of motion for a uniformly accelerating observer. The trajectory becomes 
\begin{equation}
\label{x(t)}
x(t) = x_0 \left[ 1 + \left( \frac{v_\infty t}{x_0} \right)^2 \right]^{1/2}
\end{equation}
For large $t$, we see that (\ref{x(t)}) reduces to the equation of a line $x(t)= v_\infty t$ with slope $v_\infty$. This explicitly shows that the hyperbolas cross over the 45-degree lines that define event horizons for uniformly accelerating observers. The same conclusion follows from the velocity
\begin{equation}
\label{v(t)}
v(t) = \frac{v_\infty^2 t}{x_0} \left[ 1 + \left(\frac{v_\infty t}{x_0}\right)^2 \right]^{-1/2}
\end{equation}
by considering again large $t$ behavior, where $v$ becomes $v_\infty$.

The acceleration is deceptively similar to that of a uniformly accelerating observer. 
\begin{equation}
a(t) = \frac{v_\infty^2 }{x_0} \left[ 1 + \left(\frac{v_\infty t}{x_0}\right)^2 \right]^{-3/2}
\end{equation}
We recover the exact answer for the acceleration of a uniformly accelerating observer by taking the limit as $v_\infty \rightarrow 1$ with $x_0$ (equivalently, $\alpha \gamma_{\infty}$) fixed. The proper acceleration is 
\begin{equation}
\label{a_pr(t)}
a_{pr}(t) = \frac{v_\infty^2}{x_0}  \left[ 1 + \left(\frac{v_\infty t}{x_0 \gamma_{\infty}}\right)^2 \right]^{-3/2}  = \frac{v_\infty}{\alpha \gamma_{\infty}}  \left[ 1 + \left(\frac{t}{\alpha \gamma_{\infty}^2}\right)^2 \right]^{-3/2}
\end{equation}
which clearly describes a non-uniformly accelerating observer. We see exactly what happens in the case of uniform acceleration in equation (\ref{a_pr(t)}) by taking the same limit $v_\infty \rightarrow 1$ with $x_0$ fixed. The proper acceleration reduces to the constant $a_{pr} = 1/ x_0$ one would expect for a uniformly accelerating observer because the $\gamma_\infty$ dividing the time coordinate becomes infinite as we take the limit. The transformation $a_{pr} = \gamma^3 a$ from acceleration to proper acceleration is equivalent to dividing the time coordinate $t$ by a single power of $\gamma_\infty$ due to the fact that the gamma factor resulting from (\ref{v(t)}) is given by
\begin{equation}
\label{gamma(t)}
\gamma(t) = \left[ \frac{1 + \left(v_\infty t / x_0 \right)^2}{1 + \left(v_\infty t / x_0 \gamma_{\infty}\right)^2} \right]^{1/2}
\end{equation}
 where we recover the exact $\gamma$ factor for the uniformly accelerating observer in the limit $v_\infty \rightarrow 1$ as $\gamma(t) = \left[ 1 + \left(v_\infty t / x_0 \right)^2 \right]^{1/2}$.

\section{Light Cone Coordinates}

We denote by $u = t - x$ and $v = t + x$ the standard light cone coordinates in Minkowski spacetime. We define $U = T - X$ and $V = T + X$ and use hyperbolic identities to obtain the following.
\begin{equation}
u = \alpha \sinh \left( \frac{U}{\alpha} \right) \quad  ,  \quad v = \alpha \sinh \left( \frac{V}{\alpha} \right)
\end{equation}
These coordinates are completely invertible since the hyperbolic sine inverse is a well defined function on the entire domain. This isolates the coordinates, leaving $u$ as a function only of $U$ and $v$ as a function of only $V$. Note that $u = 0 \iff U = 0$ and $v = 0 \iff V = 0$, and that both coordinate systems cover the entire spacetime. In light cone coordinates the metric becomes
\begin{equation}
ds^2 = \cosh\left(\frac{U}{\alpha} \right)\cosh \left(\frac{V}{\alpha} \right) dUdV
\end{equation}

These results will be crucial in determining whether non-uniformly accelerating observers can associate thermal properties with the Minkowski vacuum \cite{DM}.

\section{Conclusions}

We now have a metric that describes observers moving with non-constant proper acceleration. Our solution reduces to the Rindler case in the limit $v_\infty \rightarrow 1$ with $x_0$ fixed. For  $v_\infty \neq 1$, the proper acceleration vanishes as $| t | \rightarrow \infty$ and is always below the Rindler value $a_{pr} = 1/ x_0$ for finite times. These observers asymptotically approach speeds less than the speed of light, characterized by $v_\infty$, and therefore do not see event horizons. The implications of our results on the quantum theory of a scalar field in non-inertial reference frames will be addressed in a separate paper  \cite{DM}.

\newpage
\section*{Figures}

\begin{figure}[h]
\begin{center}
\includegraphics[width=4.8in]{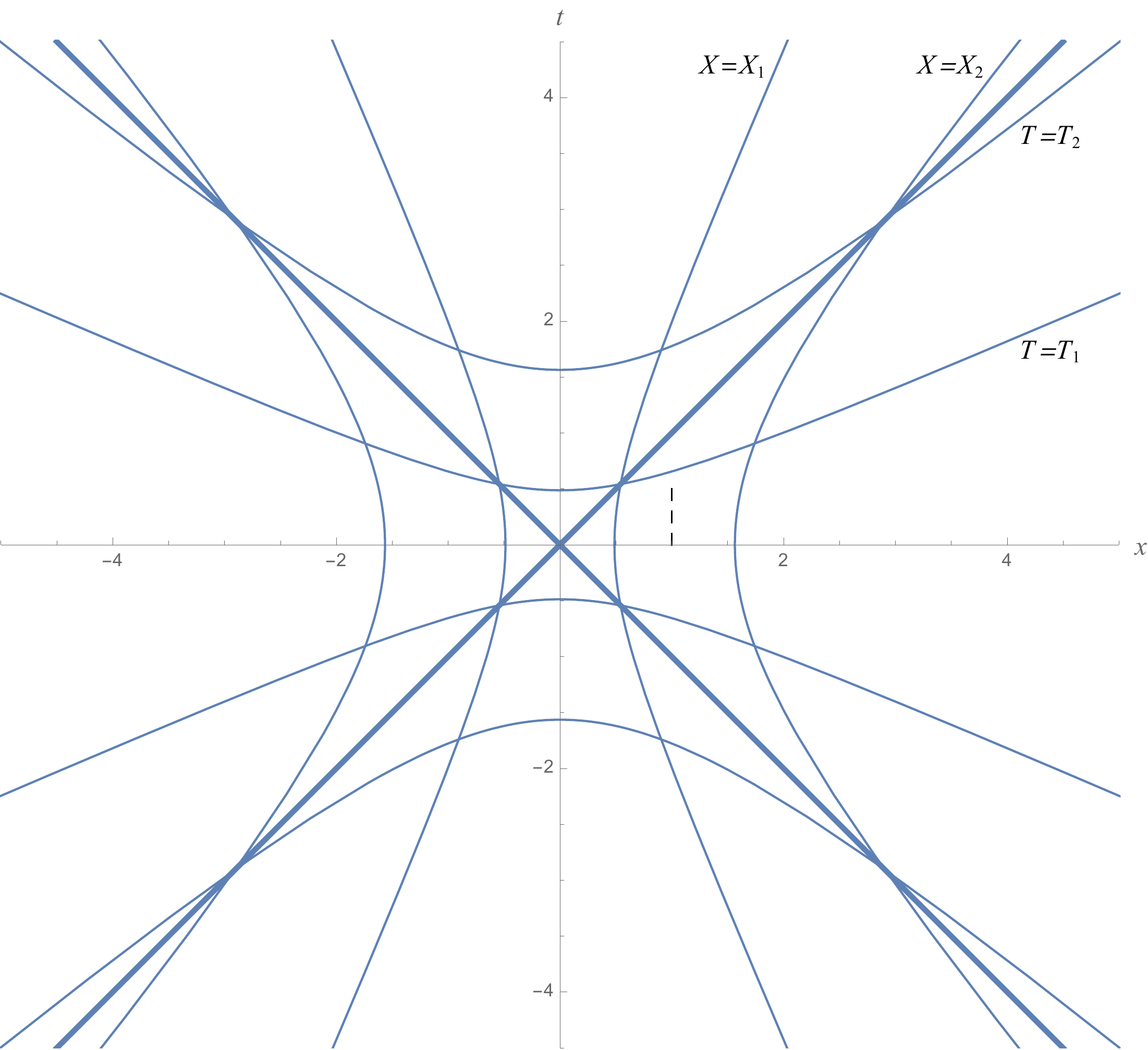}
\caption{\label{XTCoord}Spacetime diagram showing lines of constant $X$ and constant $T$. The absence of horizons for observers with constant $X$ worldines is clear from the figure.}
\end{center}
\end{figure}

\end{document}